\documentclass[a4paper]{llncs}
\usepackage{graphicx}
\usepackage{epstopdf}
\usepackage{subcaption}
\usepackage{cite}
\usepackage{xcolor}

\begin{document}
\title{DNN Based Beam Selection in mmW Heterogeneous Networks\thanks{Supported by the EXTRANGE4G project with company ETELM funded by DGA. This work has been performed at LINCS laboratory.}
}
%
%
\author{Deepa Jagyasi\orcidID{0000-0001-5592-8381} \and
Marceau Coupechoux\orcidID{0000-0003-0744-319X}}
\authorrunning{D. Jagyasi et al.}
%
\institute{LTCI, Telecom Paris, Institut Polytechnique de Paris\\
\email{\{deepa.jagyasi,marceau.coupechoux\}@telecom-paris.fr}}
\maketitle              
\begin{abstract}

We consider a heterogeneous cellular network wherein multiple small cell millimeter wave (mmW) base stations (BSs) coexist with legacy sub-6GHz macro BSs. In the mmW band, small cells use multiple narrow beams to ensure sufficient coverage and User Equipments (UEs) have to select the best small cell and the best beam in order to access the network. This process usually based on exhaustive search may introduce unacceptable latency. In order to address this issue, we rely on the sub-6GHz macro BS support and propose a deep neural network (DNN) architecture that utilizes basic components from the Channel State Information (CSI) of sub-6GHz network as input features. The output of the DNN is the mmW BS and beam selection that can provide the best communication performance. In the set of features, we avoid using the UE location, which may not be readily available for every device. We formulate a mmW BS selection and beam selection problem as a classification and regression problem respectively and propose a joint solution using a branched neural network. The numerical comparison with the conventional exhaustive search results shows that the proposed design demonstrate better performance than exhaustive search in terms of latency with at least 85\% accuracy. 


\end{abstract}

\section{Introduction}

Millimeter Wave (mmW) communication is considered as a promising technique to solve the unprecedented challenge of increasing demand for high data rates in future cellular networks. However, it suffers from limited coverage and in the ultra-dense environment it is significantly prone to blockages such as high density objects like walls, glass, humans, etc. Thus, in-order to provide flexible coverage and minimize the infrastructural cost, it is proposed that mmW networks will be deployed in a multi-tier heterogeneous network, where multiple small cell mmW base stations (BSs) coexist with multiple legacy sub-6GHz macro BSs~\cite{mmWave_network_architecture}. The legacy network operating in sub-6GHz  frequencies can handle operations like resource allocation, mobile data offloading, control signalling etc., while the potential mmW BSs can handle massive data traffic \cite{mmWave_network_architecture,mmW_future_cellular}. In this paper, we propose a solution for optimal resource allocation for a heterogeneous cellular network that enables reliable communication while leveraging the benefits of high data rates from mmW bands.

Beamforming is important in mmW systems in order to overcome the path loss due to shorter wavelength. With the large number of antenna elements associated with mmW transceivers, multiple beams are possible, which can perform directional beamforming and achieve high gain. Thus to ensure high performance, choosing the suitable BS to user equipment (UE) beam-pair from the set of all the possible directional beams is a crucial task. Beam selection has been conventionally addressed using exhaustive search or multi-level selection approach as in \cite{initial_access,two_stage_beam_selection}. However, with these techniques, large number of beams at mmW BSs leads to large beam training overhead and hence unacceptable latency to access the mmW network. Access latency is in turn significantly lower in case of communication at sub-6GHz frequencies. To overcome this challenge, out-of-band spatial information has been used for reducing beam-selection overhead \cite{out_of_band}. In recent years, in order to predict the optimal beam and significantly overcome the training overhead, the use of deep learning (DL) and machine learning (ML) tools has proved to be very promising in establishing mmW links~\cite{blockage_prediction_sub_6}. In this paper, we thus propose a deep neural network (DNN)-based mmW BS and beam selection for heterogeneous network by utilizing basic features from the Channel State Information (CSI) available only at sub-6GHz BSs.

DL and ML techniques have been hugely explored for various communication applications which include, channel estimation, design of auto-encoders, spectrum allocation, etc. \cite{ML_applications}. In the context of mmW communications, such techniques have been reported for applications such as beam selection, blockage detection, channel estimation, or proactive handover. Various DL and ML techniques to reduce the beam selection overhead in mmW communications use location information, channel information, out-of-band information or measurements from different sensors such as LIDAR, camera, or GPS. Specifically, authors in \cite{LIDAR} and \cite{LIDAR_and_position} have proposed the use of deep convolutional neural networks to perform beam selection task in distributed and centralized architecture respectively. In \cite{situational_awareness}, authors have considered the use of situational knowledge about the environment and location of UEs and proposed the use of ensemble learning-based classification to identify the optimal mmW beam. Later, in \cite{data_driven}, authors proposed the applicability of deep learning techniques such as k-nearest neighbours (KNN), support vector classifier (SVC) and multi-layer perceptron by using angle of arrival information to perform the beam-selection task. All these works however assume single-layer networks and ignore the macro-layer of sub-6GHz BSs that will be required for a continuous connectivity.   
Only two references are dealing with ML/DL-based beam selection in  heterogeneous networks \cite{blockage_prediction_sub_6,handover_sub_6}. Authors in~\cite{handover_sub_6} have considered the CSI over sub-6GHz and kernel-based ML algorithms to assist handovers for target vehicle discovery problem and overcome coverage blindness. In~\cite{blockage_prediction_sub_6}, authors have proposed the use of sub-6GHz channel and location information for performing the beam-selection and blockage prediction task. However, the solution in \cite{blockage_prediction_sub_6} is limited to a single BS - single UE communication scenario, where the BS employs co-located sub-6GHz and mmW transceivers. 
In this paper, we extend the work done in~\cite{blockage_prediction_sub_6} by considering multiple coordinating sub-6GHz and mmW BSs to perform resource allocation for each UE in the network.

Furthermore, location is an important feature that independently can be utilized to perform the task. Most of the previously discussed work on beam selection including \cite{blockage_prediction_sub_6}, considers the availability of the UE location. However, this information may not always be readily available for many cellular devices. Also, location sensors usually has low accuracy and can result in incorrect outputs\cite{location_accuracy}. Hence, we aim to intentionally eliminate the availability of location information from the set of input features and design the proposed DNN based BS and beam selection framework for a heterogeneous mmW network.

The main contributions of this paper are listed as follows:
\begin{enumerate}
    \item To guarantee reliable communication and enhanced coverage in mmW communication, we consider the heterogeneous architecture and propose DNN-based BS and beam selection by leveraging basic signal components extracted from the sub-6GHz channel as the input features. We consider multiple coordinated sub-6GHz BSs for optimal mmW resource allocation in order to serve any UE in the network.  
    \item We propose a branched DNN-structure, which divides the problem into two sub-problems of BS selection and beam selection and is well-adapted for this application.
    \item We eliminate the use of location information from the set of input features to perform the considered task. The feature vector considered as input to the network include:  the  azimuth  and  elevation angle of arrival (AoA) from the BS, the receive signal power, the signal phase and the propagation delay.  
\end{enumerate}

The remainder of this paper is organized as follows. Sec.~\ref{system_model} describes the network and transceiver model. The proposed problem is formulated in Sec.~\ref{prob_formulation} and then the deep neural network model is discussed in Sec.~\ref{DNN_description}. Sec.~\ref{simulation_results} presents the simulation environment and performance evaluation and finally Sec.~\ref{conclusion} concludes the paper. 

\noindent {\em Notations}: Throughout this paper, we use bold-faced lowercase letters to denote column vectors and bold-faced uppercase letters to denote matrices. For any matrix ${\bf X}$, ${\bf X}^{T}$ denotes the transpose operation.

\section{System Model}
\label{system_model}
\begin{figure}[t]
\centering
\includegraphics[width=\linewidth,trim={0in 2.28in 0.5in 0.5in},clip]{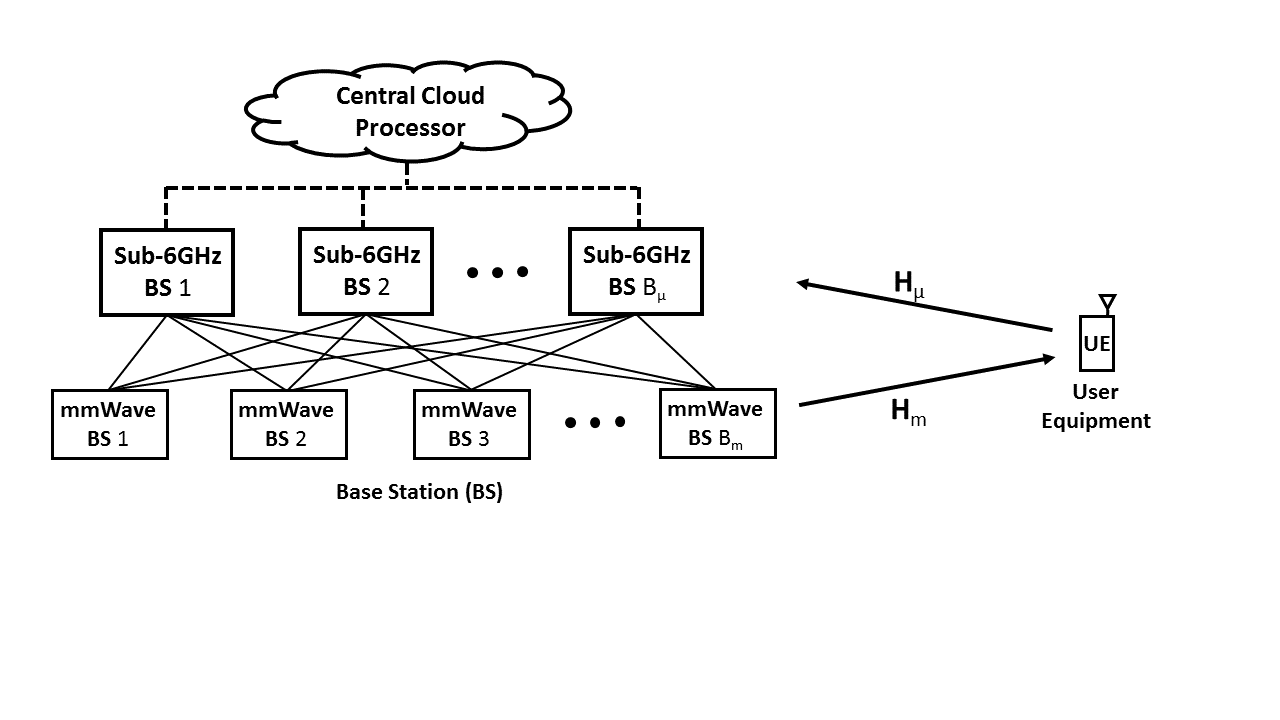}
\caption{System model: Heterogeneous network architecture with mmW small cells coexisting with sub-6GHz macro BSs. Dashed lines represent the connection of coordinating sub-6GHz BSs with a central cloud processor whereas solid lines represent the connection between any sub-6GHz and mmW BS in a network.}
\vspace{-15 pt}
\label{system_dia}
\end{figure}
We consider a heterogeneous cellular network wherein multiple sub-6GHz BSs and mmW BSs operate together in order to serve UEs in the network as shown in the Fig.~\ref{system_dia}. We assume that there are $B_{\mu}$ sub-6GHz BSs, each equipped with $N_{\mu}$ antenna elements. All the sub-6GHz BSs operate in a coordinated manner for their processing such as channel estimation or precoder design, along with DNN computations being performed at a central cloud processor unit. We assume that there are $B_{m}$ mmW BSs distributed in the network region that are coordinated with the sub-6GHz BSs to provide high speed data transfer to the UEs in the network. Each of the mmW BS is assumed to be equipped with $N_{m}$ transceiver antennas. We assume that UEs have a single antenna in both bands\footnote{UEs may be equipped with several antennas but we don't address in this paper the beam alignment problem and we focus on the beam selection at the BS. Once the BS beam is known, the UE may for example perform exhaustive search to select its own beam.}. 

The communication scenario that we study is as follows. A UE is initially connected to a sub-6GHz BS and periodically transmits pilot signals to all macro BSs. Whenever the UE is approaching towards mmW BSs, the coordinated sub-6GHz BSs command the best mmW BS and the best beam that maximizes the achievable rate for this user. 

Based on this scenario, the signal received by the macro sub-6GHz BSs at the $k$-th OFDM sub-carrier, $k=1,2,\cdots,K$ can be given by:
\begin{eqnarray}
{\bf y}_{\mu}[k]&=& {\bf h}_{\mu}[k]d_s+{\bf n}_{\mu}[k],
\end{eqnarray}
where $d_s$ is the uplink pilot transmitted over the ${\bf h}_{\mu}$ sub-6GHz channel gain matrix and ${\bf n}_{\mu}$ is the additive Gaussian noise vector with zero-mean and covariance matrix $\sigma_{\mu}^{2}{\bf I}$ at the sub-6GHz BS antenna arrays. The processing at the sub-6Ghz is performed in the baseband domain as the macro BSs are assumed to employ fully-digital architecture. 

However, due to the high cost and power consumption of mixed signal RF components at mmW frequencies, mmW transceivers are assumed to employ either fully-analog architecture where the transceiving unit is associated with single RF chain or it employs hybrid analog-digital architecture with a number of RF chains less than $N_{m}$. In this work, mmW BSs adopt fully-analog beamforming architecture where, at a given time instant, the signal is transmitted through a single beam 
which is selected from a finite set $\mathcal V$ of $M$ predefined beams, where $\mathcal V$ is the codebook.
The total transmit power at the mmW BS is $P_T$. Thus for the downlink transmission, where the mmW BS communicates with the UE, the signal received at the UE can be given as:
\begin{eqnarray}
{\bf y}_{m}[k]&=& {\bf H}_{m}[k]{\bf v}_{m}[k]d_m+{\bf n}_{m}[k],
\end{eqnarray}
where ${\bf H}_{m}$ is the mmW channel gain matrix, ${\bf v}_{m}$ is the beamforming vector,  $d_m$ is the data transmitted by the mmW BS and ${\bf n}_{m}$ is the additive Gaussian noise at UE with zero-mean and covariance matrix $\sigma_{m}^{2} {\bf I}$.

We assume that the mmW channel is modelled as a geometric channel~\cite{deep_MIMO_dataset} which can be given as: 
\begin{eqnarray}
{\bf H}_m[k]=\sum_{l=1}^{L}{\sqrt{{\frac{\rho_l}{K}}}e^{{j}(\kappa_l+{\frac{2\pi k}{K}\Gamma_l B_m})}}{\bf a}(\theta_{l},\phi_l)
\end{eqnarray}
where $\sqrt{\frac{\rho_l}{K}}$ is the path gain for the $l$-th channel path in the $k$-th OFDM sub-carrier and $\kappa_l$ and $\Gamma_l$ represents the path phase and propagation delay for the $l$-th channel path respectively. $L$ is the total number of channels paths. The array response vector at the BS is denoted by ${\bf a}(\theta_{l},\phi_l)$, where $\theta_{l}$ and \({\phi}_{l}\) is the azimuthal and the elevation AoA respectively. The detailed study of the utilized channel model can be obtained in \cite{deep_MIMO_dataset}. The sub-6GHz channel is modelled in the same way. 

\section{Problem Formulation}
\label{prob_formulation}

Given the uplink channel information at sub-6GHz BSs, we aim at designing an optimal mmW BS and beam predictor such that it maximizes the achievable sum-rate for each user in the network. Thus the optimal beamforming vector ${\bf v}_{m}^{o}$ can be obtained as:
\begin{eqnarray}
{\bf v}_{m}^{o}= {\rm arg} \max_{{\bf v}_{m}\in{\mathcal V}} \sum_{k=1}^{K} \log_2{(1+\gamma|{\bf H}_{m}[k]^{T}{\bf v}_{m}|^2)}
\end{eqnarray}
where $\gamma=P_T/{K \sigma_{m}^2}$. To design this optimal predictor we aim to find a mapping from sub-6GHz channel to mmW BS and beam selection. \cite{blockage_prediction_sub_6} has shown that, under the assumption that there is a bijective mapping between sub-6GHz channel and user location, there also exists a bijective mapping between sub-6GHz channel and mmW channel. Motivated by this result, we rely on sub-6GHz channel features to deduce the resources in mmW band. We can thus define two mapping functions $\zeta_{BS}$, $\zeta_{b}$ as follows: 
\begin{eqnarray}
\zeta_{BS}:&&{{\bf f}_{\mu}}\rightarrow {{\mathcal P}_{BS}} \\
\zeta_{b}:&& {{\bf f}_{\mu}}\rightarrow {{\bf r}_{b}}
\end{eqnarray}
where ${\bf f}_{\mu}$ is a feature vector of size $n_f$ extracted from the CSI in the sub-6GHz band, ${{\mathcal P}_{BS}}$ is a probability mass function on the set of mmW BSs and ${\bf r}_{b}$ is a vector of achievable rates for every possible beam out of $M$ at a mmW BS. To find this mapping, we utilize the DNN-based approach which are well-suited for obtaining the non-linear relationships between different data distributions\cite{deep_learning}. The $\zeta_{BS}$ mapping is formulated as a classification problem, in which each input feature is mapped into a finite set of labels; each label representing candidate mmW BSs, while $\zeta_{b}$ mapping is obtained by solving this sub-problem as a regression task wherein, a real valued achievable rate is obtained for each beam for the selected BS from $\zeta_{BS}$ mapping. The proposed DNN based solution is presented in details in Sec.\ref{DNN_description}.


\section{Deep Neural Network Model}
\label{DNN_description}

\begin{figure}[t]
\centering
\includegraphics[width=\linewidth,trim={0in 0.5in 0.3in 0in},clip]{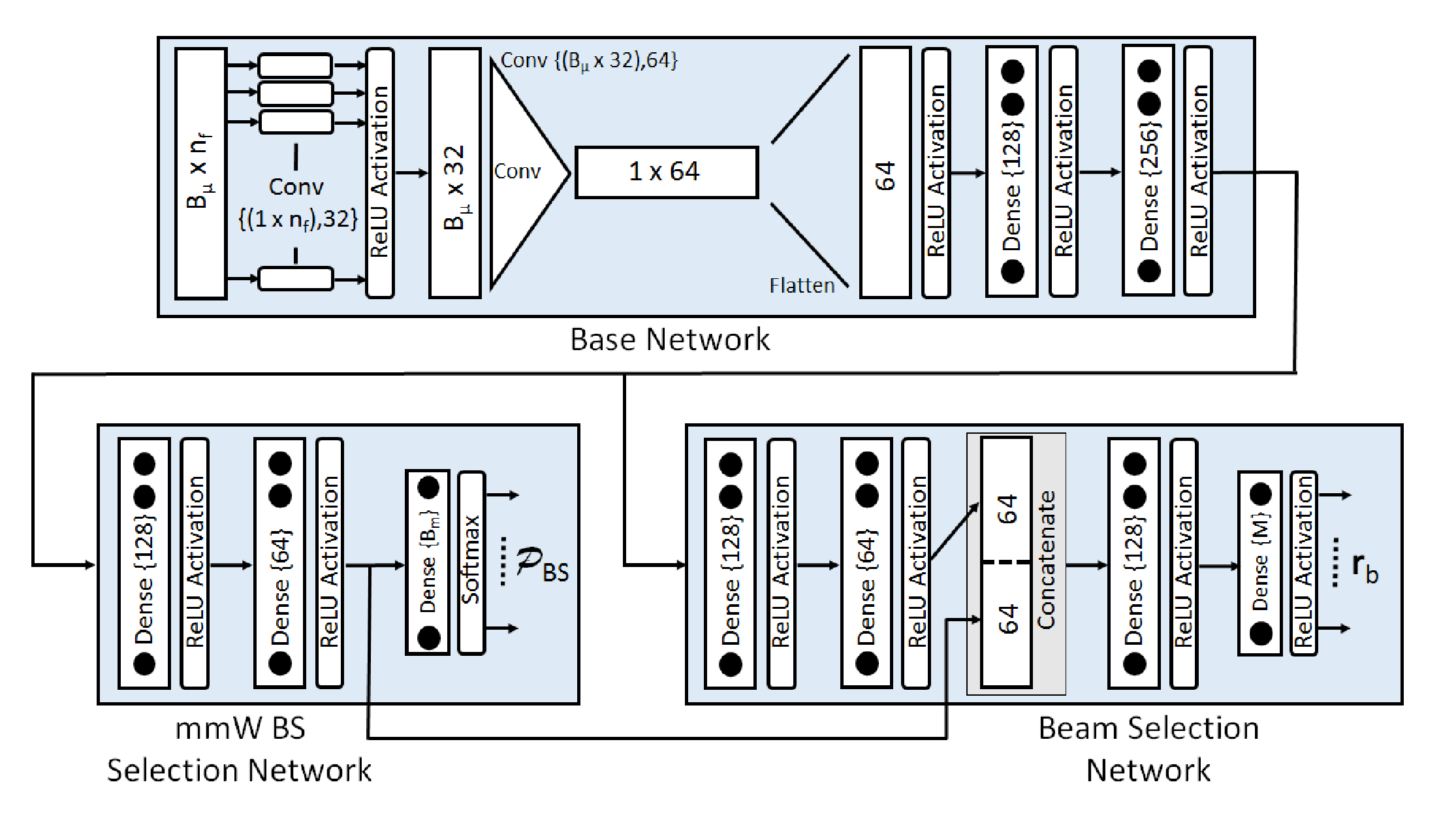}
\caption {Deep Neural Network (DNN) model for optimal mmW BS and beam selection.}
\vspace{-16 pt}
\label{DNN_model}
\end{figure}

In this section, we discuss the DL model adopted to learn the mapping from sub-6GHz channel information to mmW-BS identifier and its beam for a given user. In an environment with multiple mmW BSs and large number of beams, it is important to have flexibility in the network to incorporate new BSs or beams for future requirements. To allow this scalability, the overall beam selection problem can be divided into two sequential sub-problems: optimal mmW BS selection and then optimal beam selection. 
In order to incorporate the two sub-problems in a single neural network, we consider a branched network which takes feature vectors ${\bf f}_{\mu}$ from sub-6GHz CSI as input and predicts both mmW BS and beams for that user as shown in Fig~\ref{DNN_model}.


\subsection{Base Network}
We consider a base network for both the sub-problems to learn the common feature vectors. The input of this base network is a matrix of dimension $B_\mu \times n_f$ which gathers all the features for every sub-6GHz BS. We consider a convolution layer as the first layer of the base network with the kernel of size \(1 \times n_f\). This layer acts as a shared weight perceptron layer which is intended to find the correlation within the feature vector of each coordinating sub-6GHz BS. The output of this layer is passed through another convolution layer having kernel size $B_{\mu} \times 32$. The second convolution layer is intended to learn the correlation between the different macro BSs. We then flatten the output and pass the learned features through a stack of two fully-connected dense layers of size 128 and 256 respectively. All the layers are with Rectified Linear Unit (ReLU) non-linearity activation function as in Fig.~\ref{DNN_model}. 
The output of the final layer of the base network is branched into two sub-networks that are designed to solve each of the sub-problem of mmW BS and beam selection as discussed in following subsections.

\subsection{mmW BS Selection Network}
This sub-network is designed to predict the optimal mmW BS in order to serve the desired UE in the communication area. The input to this network are the features learned from the base network. This input vector is further passed through two fully-connected dense layers of size 128 and 64 respectively, for the optimal BS selection specific feature learning. These learned features are then projected onto the $B_{m}$ feature space using a final dense layer. The output of this layer is then fed to a softmax activation which results in a probability distribution over the number of mmW BSs. The BS with the highest probability is selected as the optimal BS. 

\subsection{Beam Selection Network}
The beam selection sub-network utilizes the learned features from the base network in order to predict the best beam. We incorporate two fully-connected dense layers of size 128 and 64 respectively, each of which is followed by ReLU activation. Moreover as selection of the best beam also depends or gets impacted by the selected BS, we concatenate the feature from the hidden layer of the BS selection network with the output of the previous dense layer from this network as depicted in Fig.~\ref{DNN_model}. These concatenated features provide added information and hence result in better performance. The output of this layer is further passed through a fully-connected dense layer of size 128, to learn the correlation within the concatenated features. Finally, we project these learned features to \(M\) dimensional space and pass it though a ReLU activation layer to get the regression output for the achievable sum-rate at each beam. The index with maximum sum-rate value is the selected beam for the selected BS from the 
mmW BS selection network. 

\subsection{Discussions}

The proposed branched neural network architecture has been obtained after experimenting several DNN configurations. 
In this section, we discuss these experimented models and  provide reasons for adapted changes in the final DNN model. 
We initially considered a multi-layer sequential DNN with single output vector. We took a concatenated vector of features from all sub-6GHz BSs as input and expected a single vector of achievable rates for each beam at each mmW BS as an output. Though this network architecture is simple and performs the task directly, we observed that this network show large variations for small changes in the environment. Moreover, when the number of mmW BSs increases, the number of output nodes increases dramatically and the system thus requires extensive training to achieve good performance.

To overcome this issue, we adopted a branched network, where we separately selected the optimal mmW BS and then the optimal beam by solving both mappings as a classification problem. Branching the complete problem to two sub-problems helped in the learning of the system and also showcased small variations for small changes in the environment. The consideration of BS selection as a classification problem performed well. It was however much less efficient for beam selection. The reason lies in the fact that due to the large number of narrow beams at mmW BSs, the angular difference between any two adjacent beams is very small, implying that multiple beams can be selected as best beam for certain user locations. We observed that this overlapping beam behaviour could not be solved by classification and the network was unable to converge to a solution. 

To tackle this issue, we modified our branched network where this time, we considered the beam selection as a regression problem. To further improve the performance of the overall system, we formed a link between the BS selection branch and the beam selection branch as both of these operations are not mutually independent. 

We adopt a soft decision for the BS selection process, i.e., we compute for every BS the selection probability and retain the one with the highest probability. In contrast, a hard decision would have selected a BS with a probability higher than a certain and given threshold. Hard decision has been observed to be training data centered and can guarantee to provide good solutions for features within the bounded range of the training data. However, a hard decision may fail to give good solutions for feature values outside these bounds. A soft decision however, will still provide a solution. Also, when all the BSs are equally probable for selection, a hard decision threshold greater than $1/B_{m}$ will not provide any solution, while a soft decision will select any one of them. Furthermore, beam selection task is also modelled with soft decision, where the best beam is selected as the one with highest achievable sum-rate. This allows for multiple beams selection (by considering the first highest sum-rates), a characteristic we will use to improve the accuracy of the results, as shown in the next section. 



\section{Simulation Results and Evaluation}
\label{simulation_results}
In this section, we illustrate the performance of the proposed DNN based BS and beam selection in a heterogeneous mmW networks. We first describe the setting of a simulation environment considered throughout the simulations in subsection~\ref{sim_env} and then discuss the performance results in subsection~\ref{results}. 
\subsection{Simulation Environment}
\label{sim_env}
We consider the outdoor simulation environment provided with the available open source DeepMIMO dataset~\cite{deep_MIMO_dataset}. From the dataset, we consider two different ray tracing scenarios 'O1\_3p5' and 'O1\_28' operating at 3.5 GHz and 28 GHz frequencies respectively, in order to construct a heterogeneous simulation environment. We consider two sub-6GHz coordinated BSs and eight mmW BSs. The deepMIMO dataset generates the channel at these frequencies. Given the CSI, we extract the basic components and construct the feature vectors from utilizing only the sub-6GHz channel, which acts as the input to our proposed DNN model. Essentially, we consider the azimuthal and elevation AoA, signal power, path loss and signal phase as the extracted features from the sub-6GHz CSI. Intentionally, we don't assume the availability of the UE location, as this information may not be available at the device. The hyperparameters considered for the generation of the dataset for training and testing are given in Table~\ref{parameters}. 
\begin{figure}[t]
\centering
\includegraphics[width=0.65\linewidth,trim={0.8in 0in 0.6in 0.4in},clip]{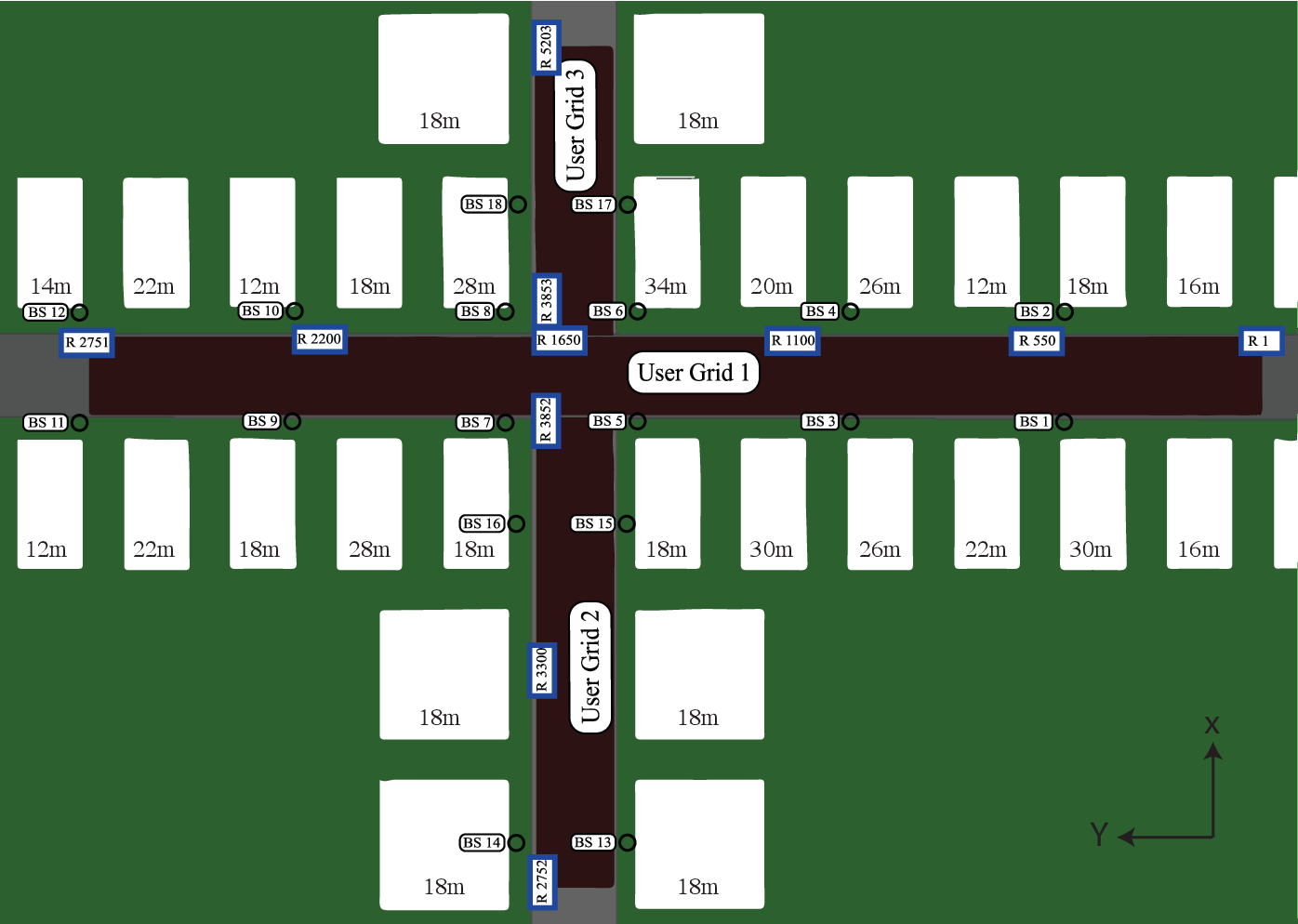}
\caption{Simulation Environment \cite{deep_MIMO_dataset}.}
\vspace{-10 pt}
\label{fig_sim}
\end{figure}
The outdoor simulation environment we considered is given in Fig.~\ref{fig_sim}. It is an urban environment with the BSs placed along the side of the road. We considered a subset of BSs and users for our experiments, the list of which is given in Table~\ref{parameters}. Users are considered to be present on the road and are densely populated for better data generation. Building of varying height, width and material are placed along the road providing blockages and reflections. For both scenarios, we considered 1024 OFDM subcarriers with an OFDM sampling factor of one, where sampling factor is the rate at which we can sample the OFDM subcarriers. Furthermore, the OFDM limit specifies the number of sampled subcarriers to be considered. We set this limit to 64 for both scenarios, which implies that we calculate the channels only at the first 64 sampled subcarriers. Detailed explanation about the simulation environment can be referred in~\cite{deep_MIMO_dataset}. 

\begin{table}[t]
\centering
\caption{Dataset parameters for mmW BSs operating at 28 GHz and macro sub-6GHz BSs operating at 3.5 GHz.}
\label{parameters}
 \begin{tabular}{||c |c |c||} 
 \hline
 Parameters & 28GHz Scenario & 3.5GHz Scenario\\ [0.5ex] 
 \hline\hline
 Active BSs & 2,3,4,5,6,7,8,17 & 1,18\\ 
 \hline
 Active users & 1651-2200, 3500-5203 & 1651-2200, 3500-5203\\
 \hline
 Number of BS Antennas & 256 & 16\\
 \hline
 Antenna spacing ($\times$wavelength) & 0.5 & 0.5\\
 \hline
 Bandwidth (GHz) & 0.5 & 0.02\\
 \hline
 Number of OFDM subcarriers & 1024 & 1024\\
 \hline
 OFDM sampling factor & 1 & 1\\
 \hline
 OFDM limit & 64 & 64\\
 \hline
 Number of paths & 1 & 1\\
 \hline
\end{tabular}
\end{table}

\subsection{Performance Evaluation}
\label{results}
In this subsection, we present the simulation results demonstrating the performance of the proposed scheme, while analyzing the effect of the number of selected beams, the training dataset selection, and the location parameter on accuracy and latency. 
\begin{figure}[t]
\centering
\includegraphics[width=0.65\linewidth,trim={0.7in 0in 0.6in 0in},clip]{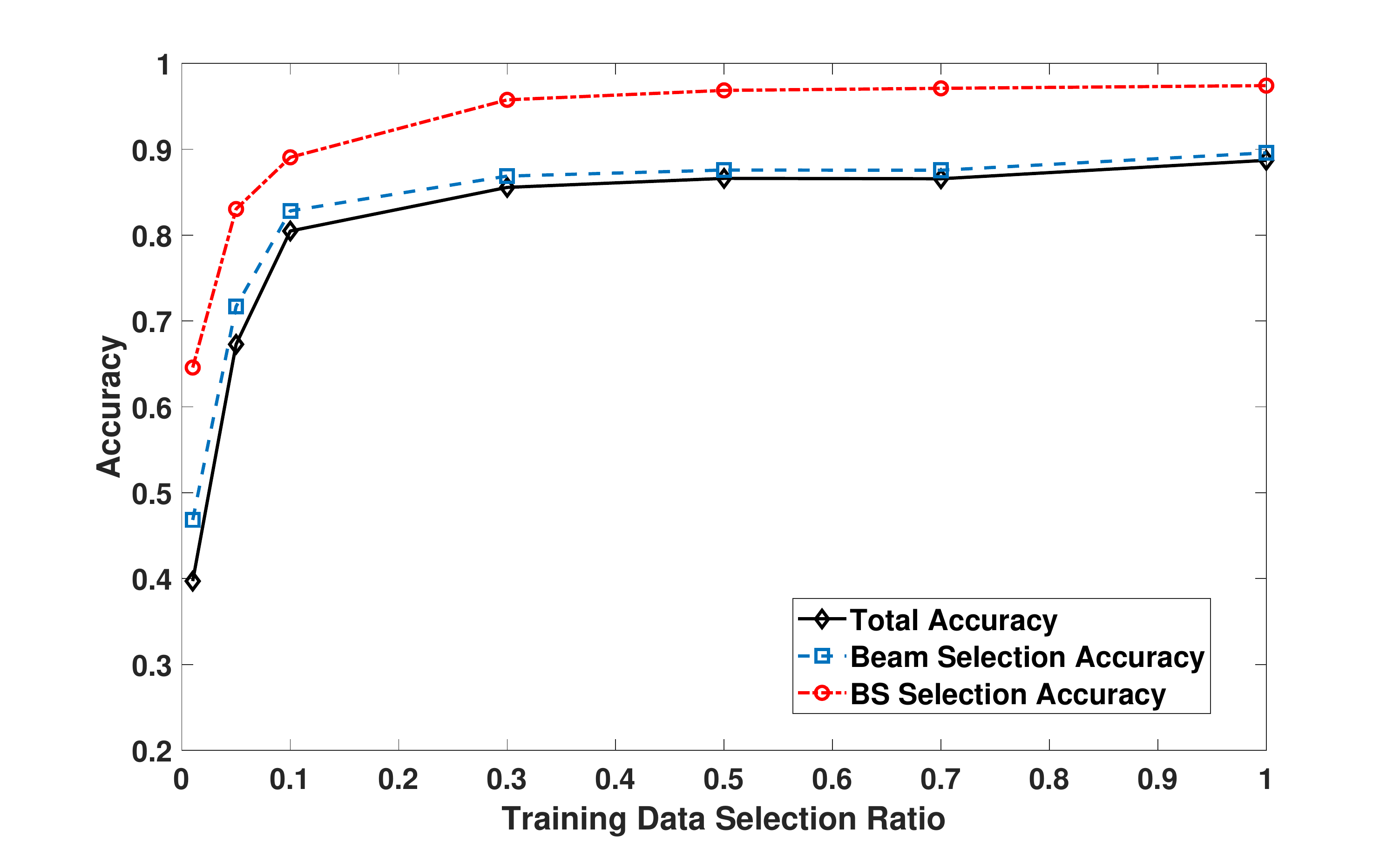}
\caption{Performance evaluation of the BS selection, beam selection and total accuracy Vs varying ratio of training data with respect to total training dataset.}
\vspace{-10 pt}
\label{prediction_acc}
\end{figure}

In Fig.~\ref{prediction_acc}, we evaluate the performance of the proposed DNN-based BS prediction, best beam prediction and overall prediction accuracy, where the total accuracy is obtained by correct prediction of both BS and beam against the varying size of the training dataset. We divide the overall data with a 80:20 ratio where 80$\%$ of the total data is used for training whereas the remaining 20$\%$ dataset is used for validation/testing purpose. Out of this total available 80$\%$ training dataset, we utilise varying training data ratios and observe the performance in terms of accuracy for the proposed system. The system performance illustrates that the network is able to achieve high accuracy for both BS and beam selection tasks. The achievable BS selection accuracy is around 97$\%$ whereas the beam can be predicted with 88$\%$. The total accuracy of correctly predicting both the optimal BS and beam is close to 86$\%$ when we use complete training dataset. We observe comparable performance with 50$\%$ of training data as compared to complete training dataset. This means that we can quickly obtain good results offline and apply the algorithm online and then improve the performance by training over the time. 

\begin{figure}[t]
\centering
\includegraphics[width=0.65\linewidth,trim={0.7in 0in 0.6in 0.4in},clip]{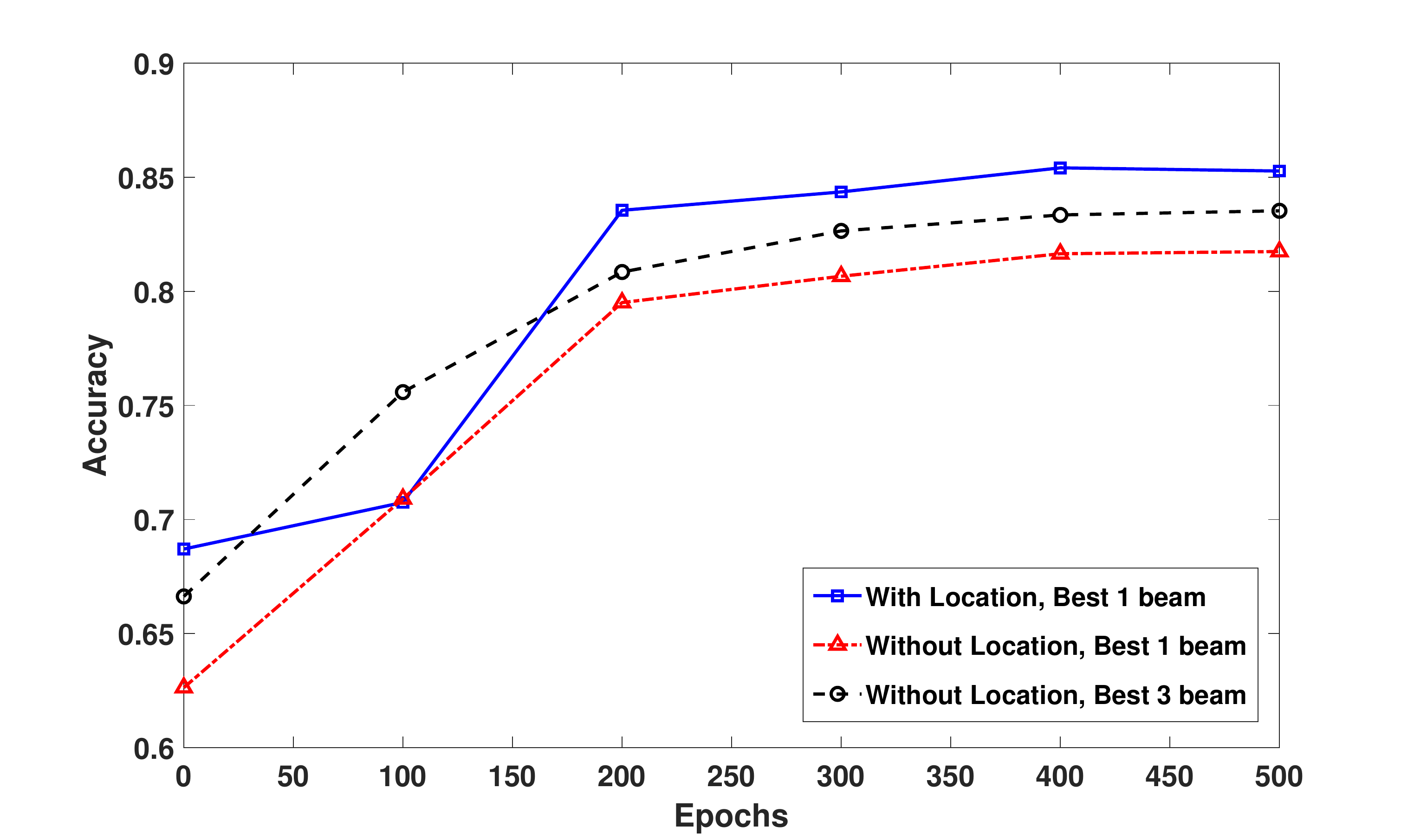}
\caption{ Accuracy Vs number of epochs comparing proposed DNN architecture based beam selection predicting best 1 beam and best 3 beams without considering location information in features and proposed DNN architecture predicting best beam while considering the location information.}
\vspace{-10 pt}
\label{fig:5a}
\end{figure}

We compare the performance of the proposed DNN architecture while now considering the UE location as one of the input features. Fig~\ref{fig:5a} shows this performance as a function of the number of epochs. We compare the performance for the best beam and the best three beams with and without location. As expected, we observe that the location-aided design performs better.
\begin{figure}[t]
\centering
\includegraphics[width=0.65\linewidth,trim={0.7in 0in 0.6in 0.4in},clip]{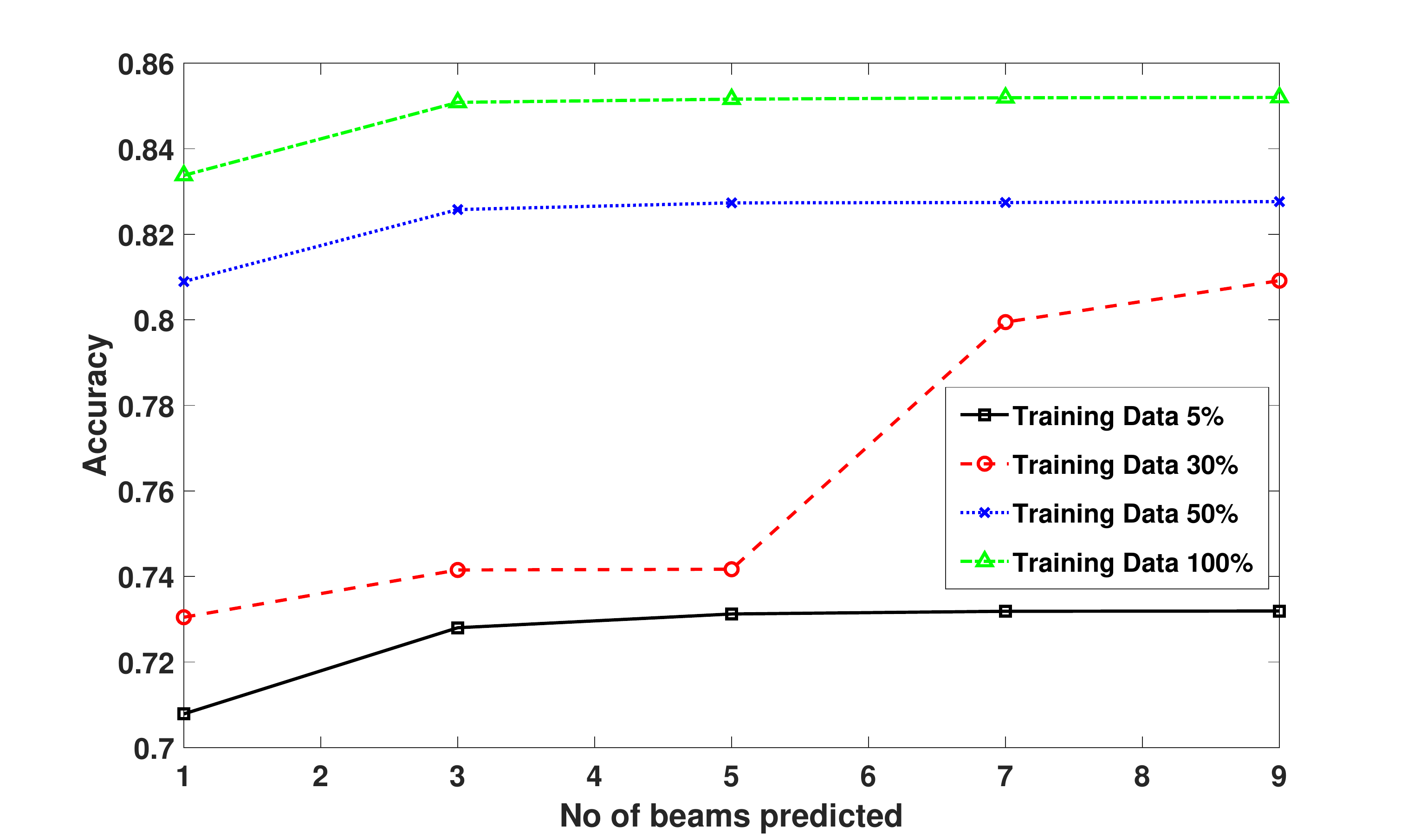}
\caption{ Accuracy Vs number of predicted beams  by the proposed DNN beam selection with varying training dataset.}
\vspace{-10 pt}
\label{fig:5b}
\end{figure}
However, the performance can be improved by selecting the best $b$ beams, $b=1...M$, hence reducing the performance gap between the architecture with or without location parameter. In Fig.~\ref{fig:5b}, we demonstrate the beam selection accuracy with respect to number of beams predicted for the proposed DNN for varying size of the training dataset. As expected, it is observed that the beam selection accuracy increases with the increasing number of predicted beams as well as with the increasing size of the training data. From this figure, we can further observe and analyse the effect of latency for the proposed system. Indeed, an exhaustive search would require to perform $M$ received power measurements (64 in our case), while with our solution, we can can achieve 85\% accuracy by measuring only the best three beams selected by the network. 
\section{Conclusion}
\label{conclusion}

 In this paper, we propose a branched DNN model that jointly performs the mmW BS prediction and beam selection task in a heterogeneous network architecture. We consider that multiple mmW BSs coexist with multiple legacy sub-6GHz BSs to serve the UE in the network area. The sub-6GHz BSs are assumed to function in a coordinated manner and are supported by the central cloud processor. We formulate the mmW BS prediction as a classification problem whereas the optimal beam selection is mapped into a regression problem. For both the tasks, we utilize the channel components available only at the sub-6GHz BSs as a set of input features. As the location information may not be always available or it can be inaccurate due to sensor errors, we intentionally eliminate the use of location as an input feature for the proposed problem. We compare the performance of the proposed DNN based design with conventional exhaustive search and observe the success probability close to 1 for allocating optimal mmW BS and beam while using reduced computational resources. Comparable performance can be achieved with and without user location available provided that the three best beams are considered. At last, we show that much fewer beam power measurements are required compared to exhaustive search, which results in lower latency.  
 

%
%
%
\bibliographystyle{splncs04}
\bibliography{ref_updated.bib}
%






\end{document}